\documentclass{pasj01}

\usepackage{color}
\usepackage{natbib}

\Received{$\langle$reception date$\rangle$}
\Accepted{$\langle$acception date$\rangle$}
\Published{$\langle$publication date$\rangle$}

\begin{document}

\title{Cloud-cloud collision in the DR 21 cloud as a trigger of massive star formation}

\author{Kazuhito \textsc{Dobashi}\altaffilmark{1}, Tomomi \textsc{Shimoikura}\altaffilmark{1},
Shou \textsc{Katakura}\altaffilmark{1},
Fumitaka \textsc{Nakamura}\altaffilmark{2,3}, and
Yoshito \textsc{Shimajiri}\altaffilmark{4}}
\altaffiltext{1}{Department of Astronomy and Earth Sciences, Tokyo Gakugei University, Koganei, Tokyo  184-8501, Japan} 
\altaffiltext{2}{National Astronomical Observatory of Japan, Mitaka, Tokyo 181-8588, Japan}
\altaffiltext{3}{Department of Astronomical Science, School of Physical Science, SOKENDAI (The Graduate University for Advanced Studies), Osawa, Mitaka, Tokyo 181-8588, Japan}
\altaffiltext{4}{Laboratoire AIM, CEA/DSM-CNRS-Universit\'e Paris Diderot,
IRFU/Service d'Astrophysique, CEA Saclay, 91191, Gif-sur-Yvette, France}
\email{dobashi@u-gakugei.ac.jp}


\maketitle


\begin{abstract}
We report on a possible cloud-cloud collision in the DR 21 region,
which we found through molecular observations with the Nobeyama 45-m telescope.
We mapped an area of $\sim 8\arcmin \times 12\arcmin$ around the region
with twenty molecular lines including the $^{12}$CO($J=1-0$) and $^{13}$CO($J=1-0$)
emission lines, and sixteen of them were significantly detected.
Based on the $^{12}$CO and $^{13}$CO data, we found
five
distinct velocity components 
in the observed region, and we call molecular gas associated with these components
``$-42$",``$-22$", ``$-3$", ``$9$", and ``$17$" km s$^{-1}$ clouds taking after their typical radial velocities.
The $-3$ km s$^{-1}$ cloud is the main filamentary cloud ($\sim31,000$ $M_\odot$) associated with
young massive stars such as DR21 and DR21(OH), and the $9$ km s$^{-1}$ cloud is a smaller
cloud ($\sim3,400$ $M_\odot$) which may be an extension of the W75 region in the north.
The other clouds are much smaller.
We found a clear anticorrelation in the distributions of
the $-3$ and $9$ km s$^{-1}$ clouds, and detected faint $^{12}$CO emission
having intermediate velocities bridging the two clouds at their intersection.
These facts strongly indicate that the two clouds are colliding against each other.
In addition, we found that DR21 and DR21(OH) are located in the periphery
of the densest part of the $9$ km s$^{-1}$ cloud, which is consistent
with results of recent numerical simulations of cloud-cloud collisions.
We therefore suggest that the $-3$ and $9$ km s$^{-1}$ clouds are colliding,
and that the collision induced the massive star formation in the DR21 cloud. 
The interaction of the $-3$ and $9$ km s$^{-1}$ clouds was previously
suggested by \citet{Dickel1978}, and our results strongly support their hypothesis
of the interaction.

\end{abstract}


\section{Introduction}\label{sec:introduction}
DR21 is a small H{\scriptsize II} region ($\sim30\arcsec \times 30\arcsec$)
in the Cyg X region and is very bright in the radio continuum
\citep{Wendker1984}
and in the infrared vibration line of molecular hydrogen tracing powerful outflow(s)
driven by extremely luminous ($\sim10^5$ $L_\odot$)
young stellar objects (YSOs) forming there \citep[e.g.,][]{Garden1986,Garden1991}.
DR21 hosts a cluster having a high star density 
\citep[$\sim1\times 10^3$ stars pc$^{-2}$, ][]{Kuhn2015} and it
presumably includes several O-type stars \citep[][]{Harris1973,Roelfsema1989,Cyganowski2003}
as well as very young members that can be classified to Class I or earlier \citep{Marston2004}.
Natal cloud of DR21 is a dense massive filament extending over $\sim15\arcmin$
toward north-south direction,
and other massive YSOs known as
DR21(OH)\citep[e.g.,][]{Norris1982}
as well as
W75S-FIR1,
W75S-FIR2,
and
W75S-FIR3 \citep[][]{Harvey1986}
are associated with the cloud.

A number of observations have been conducted toward this cloud.
Dust continuum observations have unveiled the global structure of the cloud \citep[e.g.,][]{Hennemann2012},
and scores of YSOs and dense cloud cores have been found along the ridge of the filamentary
cloud \citep[e.g.,][]{Kumar2006,Motte2007}. 
Extensive molecular observations in the millimeter-wavelengths have also been carried out
and various molecular lines such as CO, SiO, CH$_3$OH, and H$_2$CS have been detected
\citep[e.g.,][]{Loren1986,Kalenskii2010,Schneider2010}.
Energetic molecular outflows have been revealed and studied in details
\citep[e.g.,][]{Garden1986,Davis2007,Zapata2012,Garden2014,DuarteCabral2014,Ching2018}.
Recent parallax measurements of DR21 indicate a distance of $1.50^{+0.08}_{-0.07}$ kpc
\citep{Rygl2012} which we adopt as the distance of the DR21 cloud in this paper.

It is known that there are two main velocity components toward the DR21 cloud.
One is at $V_{\rm LSR}\simeq -3$ km s$^{-1}$ which comes
from the main cloud forming the massive YSOs.
The other is at $\sim9$ km s$^{-1}$ which may originate from the W75 region in the north.
In an early molecular study of this region, \citet{Dickel1978} suggested that the two clouds
are interacting with each other,
based on the results of their  $^{12}$CO and $^{13}$CO observations.
Such interaction (cloud-cloud collisions) should sufficiently compress the clouds and may induce
massive star formation. 
The process of cloud-cloud collisions have been long investigated theoretically
\citep[e.g.,][]{Scoville1986,Habe1992,Tan2000,Dobbs2015,DuarteCabral2011,
Matsumoto2015,Wu2017a,Wu2017b,Takahira2018}.
A clear observational evidence of the actual cloud-cloud collisions
was first discovered by \citet[][]{Hasegawa1994}
toward the Sgr B2 giant molecular cloud. 
More recently, a number of evidences of cloud-cloud collisions have been found
in many massive star forming regions
\citep[e.g.,][]{Higuchi2009,Torii2011,Nakamura2014, Fukui2018,Nishimura2018},
suggesting that the cloud-cloud collisions can be one of the major triggers of
massive star formation.

It is very possible that the massive star formation in the DR21 cloud was
also triggered by such cloud-cloud collisions.
The suggestion of \citet{Dickel1978} is important, but
their original $^{12}$CO and $^{13}$CO data were obtained in 1970's
with the NRAO 11-m telescope
and are rather poor compared to those obtained by more recent instruments.
In addition, their suggestion is based only on a positional
coincidence of the two clouds on a large scale and a velocity
gradient of the cloud at $9$ km s$^{-1}$.
More sensitive dataset as well as detailed analyses are definitely
needed to confirm and develop their hypothesis.
For this purpose, we have carried out observations toward the DR21 cloud
with various molecular lines at much higher angular resolutions
using the 45-m telescope at the Nobeyama Radio Observatory (NRO).

The observations presented in this paper were made
as a part of ``Star Formation Legacy Project" at the NRO (led by F. Nakamura) to observe
star forming regions such as
Orion A, Aquila Rift, M17, and a few other clouds such as northern coal sack (NCS). 
An overview of the project \citep{Nakamura2019a}
and detailed observational results for the individual regions
are given in other articles
(Orion A:
Nakamura et al. 2019b,
Ishii et al. 2019,
Tanabe et al. 2019,
H. Takemura et al. in preparation,
Aquila Rift:
Shimoikura et al. 2019,
Kusune et al. 2019,
M17:
T. Shimoikura et al. in preparation,
Q. Nguyen Lu'o'ng et al. in preparation,
Sugitani et al. 2019,
NCS:
Dobashi et al. 2019)

In this paper, we report results of the observations toward the DR21 cloud.
We describe the observational procedures
in section \ref{sec:observations}.
We detected five
velocity components including those at $V_{\rm LSR}\simeq -3$ and $9$
km s$^{-1}$, and find a clear anticorrelation in their spatial distributions.
These results are shown in section \ref{sec:results}.
In section \ref{sec:discussion}, we discuss the possible cloud-cloud collisions of the
two velocity components, and summarize our conclusions in section \ref{sec:conclusions}.


\section{Observations }\label{sec:observations}

Observations were carried out with the NRO 45-m telescope in the period between 2013 March and 2013 May
as well as in 2014 April. We observed twenty molecular emission lines at $86-115$ GHz
including the $^{12}$CO($J=1-0$), $^{13}$CO($J=1-0$), C$^{18}$O($J=1-0$),
HCO$^{+}(J=1-0)$, and SiO$(J=2-1, v=0)$ lines.
The observed lines are summarized in table \ref{tab:lines}.
As the front end, we used the SIS receiver named Tz \citep{Nakajima2013} which
provided a typical noise temperature of $T_{\rm sys}\simeq200$ K including the atmosphere.
As the back end, we used the SAM45 \citep{Kamazaki2012} which consists of
16 independent digital spectrometers having 4096 channels covering
the 125 MHz bandwidth with the 30 kHz resolution.
The frequency resolution corresponds to the $\sim0.1$ km s$^{-1}$ velocity resolution at $110$ GHz.

We employed On-The-Fly \citep[OTF,][]{Sawada2008} technique to map an area of $\sim 5\arcmin \times 10\arcmin$ or
$\sim 8\arcmin \times 12\arcmin$ depending on the molecular lines to cover most of the extent of the DR21 cloud.
The OFF (i.e., emission-free) position was chosen to be 
R.A.(J2000)$=$20$^{\rm h}$57$^{\rm m}$5.04$^{\rm s}$ and
Dec.(J2000)$=$39$^{\circ}$19$\arcmin$59.4$\arcsec$.
Intensity calibration was done with the standard chopper-wheel method \citep{Kutner1981},
and spectra in units of $T_{\rm a}^*$ were obtained. Further calibration to convert the data to $T_{\rm mb}$
was done by applying the beam efficiency of the 45-m telescope which changes
from $\eta=35.9$ $\%$ at $86$ GHz to $30.4$ $\%$ at $115$ GHz.
Pointing accuracy was better than $5\arcsec$, as was checked by observing the
SiO maser sources T-Cep and IRC+60334 every $\sim2$ hours during the observations

Data reduction was done with a software package named NOSTAR.
We removed the linear baselines from the spectral data, and
resampled the data onto the $10\arcsec$ grid and then smoothed them with a Gaussian kernel
to produced the spectral data cube having an angular resolution of $\sim23\arcsec$ (FWHM)
and velocity resolution of 0.1 km s$^{-1}$.
The noise levels of the final spectral data are in the range
$\Delta T_{\rm rms}=0.2-1.5$ K  in units of $T_{\rm mb}$ for the velocity resoluton of 0.1 km s$^{-1}$,
as summarized in table \ref{tab:lines}.

 
\section{Results}\label{sec:results}

\subsection{Molecular distributions} \label{sec:distributions}
As indicated in table \ref{tab:lines}, sixteen molecular lines were detected in the observed region.
Among these, we show in figure \ref{fig:iimap1} the velocity-integrated intensity maps
of six molecular lines which we use to investigate the structures of the DR21 cloud in this paper.
We show intensity maps of the other ten molecular lines in the appendix (figure \ref{fig:iimap2}).
In figure \ref{fig:iimap1},
we indicate locations of five well-known YSOs, i.e., DR21, DR21(OH), W75S-FIR1, W75S-FIR2,
and W75S-FIR3 whose coordinates are taken from \citet{Motte2007}.
Some of the emission lines such as HCO$^{+}$($J=1-0$)
have already been observed
by earlier studies \citep[e.g.,][]{Dickel1978,Schneider2010},
and their distributions are similar to
those shown in the figure.

The molecular lines are strongly detected especially toward
DR21 and DR21(OH). In figure \ref{fig:spectra}, we show some spectra observed toward
the five YSOs shown in figure \ref{fig:iimap1}
as well as toward a position labeled ``9 km s$^{-1}$ cloud" in table \ref{tab:line_parameters}
(see section \ref{sec:ccc}).
In the $^{12}$CO, $^{13}$CO, and C$^{18}$O spectra, there can be seen some well-separated,
distinct velocity components. We fitted the individual components seen in the CO lines with
a Gaussian function. In table \ref{tab:line_parameters}, we summarize the peak brightness
temperature $T_{\rm mb}$,
the centroid velocity $V_{\rm LSR}$,
and the FWHM line width $\Delta V$ best fitting the spectra.

Five velocity components can be recognized in the CO lines.
The brightest component in CO is found at $V_{\rm LSR}\simeq -3$ km s$^{-1}$
which originates from the main body of the DR21 cloud, and 
the second brightest component is seen at $V_{\rm LSR}\simeq 9$ km s$^{-1}$
which may be a part of the W75 region. These velocity components are
consistent with those found by earlier studies \citep[e.g.,][]{Dickel1978,Schneider2010}.
In our data, another velocity component is seen at $V_{\rm LSR}\simeq 17$ km s$^{-1}$
in the $^{12}$CO spectrum shown in figure \ref{fig:spectra}f.
Furthermore, weak emission is seen at $V_{\rm LSR} \simeq -42$ km s$^{-1}$ in $^{12}$CO
only around W75S-FIR1 and W75S-FIR2 (figure \ref{fig:spectra}c and \ref{fig:spectra}d ).
In addition, there is faint $^{12}$CO emission ($\lesssim 5$ K) around $-22$ km s$^{-1}$
in the north-western corner of the mapped area (figure \ref{fig:channel_maps}b).
In this paper, we will refer to molecular clouds associated with these velocity components as
``$-42$", ``$-22$", ``$-3$", ``9", and ``17" km s$^{-1}$ clouds taking after their typical radial
velocities and following the nomenclature of \citet{Dickel1978}.


\subsection{Anticorrelations and masses of the clouds}\label{sec:ccc}

In this paper, we will mainly use the $^{12}$CO and $^{13}$CO data to investigate
the distribution of the clouds as well as their possible interaction.
Figure \ref{fig:channel_maps} shows the channel maps of the $^{12}$CO and $^{13}$CO emission lines
integrated over different velocity ranges indicated on the top of each panel.
In the figure, panel (a) shows the $^{12}$CO distribution of the $-42$ km s$^{-1}$ cloud.
The $^{12}$CO emission in the north-western corner of panel (b) traces
the $-22$ km s$^{-1}$ cloud, but the emission in the middle of the panel
entirely traces a part of the blue-shifted high-velocity gas from the outflows driven
by YSOs such as DR21 (see the $^{12}$CO spectra in figure \ref{fig:spectra}).
Panels (c)--(e) and (f)--(h) show the distributions of the $^{12}$CO and $^{13}$CO
emission of the $-3$, $9$, and $17$ km s$^{-1}$ clouds, respectively.

The overall distributions of the  $-3$ and $9$ km s$^{-1}$ clouds on the plane of sky
is basically consistent with those found by \citet{Dickel1978},
but it is noteworthy that our maps in figure \ref{fig:channel_maps} clearly reveal
their anticorrelation:
The $^{13}$CO intensity peak position of the $9$ km s$^{-1}$ cloud
(figure \ref{fig:channel_maps}g) coincides with the valley of the $-3$ km s$^{-1}$
cloud (figure \ref{fig:channel_maps}f) located in the middle of DR21 and DR21(OH).
The same trend is seen in the $^{12}$CO distributions
(figures \ref{fig:channel_maps}c and \ref{fig:channel_maps}d).
This feature can be more clearly recognized in figure \ref{fig:collision}
where the $^{13}$CO and $^{12}$CO distributions of
the two clouds are superposed.
We further point out that in the northern part of the main filament
between DR21(OH) and W75S-FIR3,
there is another anticorrelation between
the $-3$ and $9$ km s$^{-1}$ clouds in the $^{12}$CO intensity distributions
(figure \ref{fig:collision}a).
Such anticorrelations can be regarded as an evidence of cloud-cloud collisions which often trigger
massive star formation as seen in other star forming regions \citep[e.g.,][]{Dobashi2014,Matsumoto2015,Dobashi2019}. 
We will further discuss this possibility in section \ref{sec:discussion}.

We estimated molecular masses of the $-3$, $9$, and $17$  km s$^{-1}$ clouds
in a standard way \citep[e.g.,][]{Shimoikura2011} using the $^{12}$CO and $^{13}$CO data.
First, assuming that the $^{12}$CO emission line is optically very thick,
we estimated $T_{\rm ex}$ the excitation temperature of each cloud from
$T_{\rm max}^{^{12}{\rm CO}}$ the maximum
brightness temperature of the $^{12}$CO spectra at each observed position as
\begin{equation}
\label{eq:Tex}
{T_{\rm ex}} = \frac{{5.532}}{{\ln \left( {\frac{{5.532}}{{T_{\rm max}^{^{12}{\rm CO}} + 0.836}} + 1} \right)}} ~~~~{\rm K}
\end{equation}
where $T_{\rm max}^{^{12}{\rm CO}}$ is in units of Kelvin.
For positions with $T_{\rm ex}<10$ K, we assumed $T_{\rm ex}=10$ K
because our assumption of the optically thick case may not be satisfied.
We then derived $ \tau(v)$ the optical depth of the $^{13}$CO emission line as
\begin{equation}
\label{eq:tau}
\tau(v)  =  - \ln \left( {1 - \frac{{T_{\rm mb}^{^{13}{\rm CO}}(v)}}{{\frac{{{\rm{5.289}}}}{{\exp \left( {5.289/{T_{\rm ex}}} \right) - 1}} - 0.8867}}} \right)\ 
\end{equation}
where $T_{\rm mb}^{^{13}{\rm CO}}$ is a function of $v$ the radial velocity
and is in units of Kelvin.
The column density of $^{13}$CO molecules, $N$($^{13}$CO), is derived from $ \tau(v)$ and $T_{\rm ex}$ as
\begin{equation}
\label{eq:N13CO}
N(^{13}{\rm CO}) = \frac{{2.513 \times 10^{14}~{T_{\rm ex}}}}{{1 - \exp \left( { - 5.289/{T_{\rm ex}}} \right)}}\int {\tau(v) dv} ~~~~{\rm cm^{-2}}~~.
\end{equation}
Finally, we calculated the column density of hydrogen molecules $N$(H$_2$) using an empirical relation
$N$(H$_2$)$/N$($^{13}$CO)$=5\times10^5$ \citep{Dickman1978}, and calculated total masses of the clouds
within the mapped region shown in figures \ref{fig:channel_maps}f--\ref{fig:channel_maps}h
assuming a mean molecular weight of $\mu=2.8$ and a distance of $d=1.5$ kpc.

As a result, we found that the $-3$, $9$, and $17$ km s$^{-1}$ clouds have a molecular mass
of $\sim31000$, $\sim3400$, and $\sim500$ $M_\odot$ within the mapped area, respectively.
The $-3$ km s$^{-1}$ cloud is the most massive among the three clouds occupying $\sim90$ \%
of the total system,
and it has much higher excitation temperature and column density
($T_{\rm ex}\simeq 70$ K and $N$(H$_2$)$\simeq 4.7 \times 10^{23}$ cm$^{-2}$ at the maximum)
than the other clouds.
We didn't derive the
masses of the $-42$ and $-22$ km s$^{-1}$ clouds, because they are
not detected in $^{13}$CO.
These results are summarized in table \ref{tab:mass}.

The relation of the two major clouds at $-3$ and $9$ km s$^{-1}$ is of particular interest,
and we briefly estimate whether they are gravitationally bound or not.
The $-3$  km s$^{-1}$ clouds have an apparent size of $\sim8\arcmin \times 2\arcmin$ 
in $^{13}$CO with a geometrical mean radius of $\sqrt{8\arcmin \times 2\arcmin }/2=2\arcmin$,
corresponding to a radius of $R=0.9$ pc at a distance of 1.5 kpc.
The virial velocity calculated as $V_{\rm vir}=\sqrt{2GM/R}$, where $M$ is the mass of the
$-3$ km s$^{-1}$ cloud ($3.1\times 10^4$ $M_\odot$) and $G$ is the gravitational constant,
is $\sim17$ km s$^{-1}$. This value is the same order as the velocity separation
of the $-3$ and $9$ km s$^{-1}$ clouds, 12 km s$^{-1}$ in the line-of-sight velocity
or $12\sqrt{3}\simeq21$ km s$^{-1}$ in 3D, and thus they could be gravitationally bound
if they are located at the same distance.

\subsection{Molecular outflows and the SiO emission}\label{sec:outflow}

In some of the emission lines observed,
we detected molecular outflows associated with YSOs embedded in the main cloud at $-3$ km s$^{-1}$.
Figure \ref{fig:outflows} displays the blue lobes (panels a--c) and red lobes (panels d--f)
of the outflows detected in the $^{12}$CO, CS, and HCO$^+$ emission lines.
Though the red lobes detected in $^{12}$CO are highly contaminated by the other velocity components
at $9$ and $17$ km s$^{-1}$, they are traced clearly in the other two molecular lines.
All of the blue lobes are well traced in the three molecular lines.
The red and blue lobes associated with DR 21 and DR 21(OH) are huge and well defined, whereas
those in the vicinity of W75S-FIR1, W75S-FIR2, and W75S-FIR3 are not separated well
and we cannot identify the definite driving source(s) at the resolution of the current observations.

The outflow lobes associated with DR21 appear to have rather simple structures in HCO$^{+}$,
extending over the length $L\simeq1$ pc
(measured at the lowest contours in figures \ref{fig:outflows}c and \ref{fig:outflows}f)
from the driving source on the plane of sky.
The characteristic velocity defined as the maximum velocity shift of the
high velocity gas from the systemic velocity is $V_{\rm char}\simeq 25$ km s$^{-1}$
(measured using the blue wing of the HCO$^{+}$ line in figure \ref{fig:hco}).
The apparent dynamical time scale of the outflow is therefore
$\tau_{\rm age}=L/V_{\rm char} \simeq 4\times 10^4$ yr, which is consistent with the results
of \citet{Garden1986} if we take into account the difference of the assumed distance;
they derived $\tau_{\rm age}=6-12 \times 10^4$ yr in the same way as we do, but they assumed a distance of 3 kpc.
However, this time scale may be the upper limit to the actual value, because the axis of the outflow
should be nearly orthogonal to the line-of-sight, as the red and blue lobes largely overlap on the plane of sky.
As illustrated in figure \ref{fig:outflow_age}, if we assume that the outflow is ejected radially
from DR21 at a constant speed of $V_{\rm outflow}$ and that the opening angle of the outflow is
roughly $\theta \simeq 60^\circ$, the dynamical time scale of the outflow can be rescaled to
$\tau_{\rm age}=L/V_{\rm outflow}=L/V_{\rm char}{\rm sin}(\theta/2) \simeq 2 \times 10^4$ yr.

There are apparent coincidences between the outflows and the SiO emission.
The gray scale and black contours in figure \ref{fig:outflows} represent the intensity of the SiO emission.
Since \citet{Mikami1992} discovered that the SiO emission traces the shocks of the molecular
outflow in LDN 1157, the emission has been recognized as a good tracer of shocks,
and it has been widely used to study molecular outflows \citep[e.g.,][]{Hirano2001,Louvet2016,Shimajiri2008,Shimajiri2009}.
Some of the outflows in the DR21 region have been observed in SiO
by earlier studies \citep[e.g.,][]{DuarteCabral2014}.

As seen in figures \ref{fig:iimap1}d and figure \ref{fig:outflows},
the SiO emission is widely distributed in the observed region, but it
is detected only in the three limited areas just around the YSOs, and the areas
nicely overlap with the extents of the blue and/or red lobes.
We therefore regard that the SiO emission detected in our observations is mostly due to
the outflows.

However, we should note that the SiO emission not originating from outflows but originating from
some other phenomena related to the formation of molecular clouds are
recently found \citep[e.g.,][]{JimenezSerra2010, Nguyen-Lu'o'ng2013}.
The emission is generated by low-velocity shocks such as by
colliding shear or mass-accretion onto dense clumps,
not by the high-velocity outflows from YSOs. Recent numerical simulations show that,
when massive clouds collapse by the self-gravity, 
striations parallel to the magnetic field naturally form \citep[e.g.,][]{Wu2017a,Wu2017b}
along which mass is fed from the surroundings to the central dense region.
\citet{DuarteCabral2014} suggested that a certain fraction of the SiO emission
in the DR21 region can be due to such low-velocity shocks.
Global gravitational collapsing motion and convergent flows have been
suggested in this region \citep[e.g.,][]{Kirby2009,Csengeri2011}.
In addition to these suggestions, we would suggest a possibility that a certain fraction
of the SiO emission around DR21
and DR21(OH) might be due to shocks by the collision of
the $-3$ and $9$ km s$^{-1}$ clouds,
though it is very difficult to assess the contribution of this effect
in the observed SiO spectra.


\section{Discussion}\label{sec:discussion}

The results shown in the previous section indicates that at least two of the
clouds at $-3$ and $9$ km s$^{-1}$ are very likely to be
interacting, probably colliding against each other.
Such cloud-cloud collisions are known to trigger formation of massive stars
and/or star clusters \citep[e.g.,][]{Hasegawa1994, Dobashi2014,Matsumoto2015},
and we suggest that formation of the massive YSOs in the DR21 region
was triggered by the collision of the $-3$ and $9$ km s$^{-1}$ clouds.
In the following, we further discuss the possibility of their collision and its relation to
star formation based on the data obtained by our observations.

We first attempt to figure out the positional relation of the two clouds
along the line of sight.
At a glance of figure \ref{fig:collision}, we have an impression that
the two clouds have collided and already passed through each other,
and that the $9$ km s$^{-1}$ cloud is now located much farther than the
$3$ km s$^{-1}$ cloud from the observer.
However, the HCO$^{+}$ spectrum observed toward DR21
exhibits a clear dip at $V_{\rm LSR}=7-10$ km s$^{-1}$
corresponding to the velocity of the  $9$ km s$^{-1}$ cloud (figure \ref{fig:hco}).
Such a dip is often seen in optically thick molecular lines like HCO$^{+}$
toward hot regions such as compact H{\scriptsize II} regions, and it is
due to the absorption by colder gas in the foreground.
Thus, the $9$ km s$^{-1}$ cloud should be located in front of the 
$-3$ km s$^{-1}$ cloud. However, if the two components have already
passed through each other and are well separated, we would not see such a dip.
We therefore propose that the two clouds are attaching, or they are just
crossing each other at present, and an amount of colder gas of the $9$ km s$^{-1}$
cloud still remains in front of the $-3$ km s$^{-1}$ cloud (see figure \ref{fig:model}).

As we estimated in section \ref{sec:outflow}, the dynamical time scale of
the outflow associated with DR21 is
$\tau_{\rm age}\simeq 2 \times 10^4$ yr.
If we assume that the YSOs and their associated outflows were formed soon after
the collision of the $-3$ and $9$ km s$^{-1}$ clouds,
they have traveled only $0.2-0.3$ pc along the line of sight
which is much smaller than the apparent width of
the filamentary $-3$ km s$^{-1}$ cloud or the apparent diameter of the round $9$ km s$^{-1}$ cloud
($\sim2\arcmin$, corresponding to $\sim1$ pc).
If this is the case, the two clouds should be still crossing each other, which supports
the positional relation discussed in the above.

We further search for an evidence of the cloud-cloud collision.
If the $-3$ and $9$ km s$^{-1}$ clouds are colliding,
we would expect that the gas at the intermediate velocities should exist
\citep[e.g., see figures 3 and 5 of][]{Habe1992}, which could be detected
in the observed molecular lines. Figure \ref{fig:pv} shows the position-velocity
diagrams taken along the cuts parallel to the right ascension and
declination axes crossing the peak position of the $9$ km s$^{-1}$ cloud.
As seen in the figure, faint $^{12}$CO emission is detected along both of the cuts,
bridging the $-3$ and $9$ km s$^{-1}$ clouds.
We suggest that the $^{12}$CO emission with the intermediate velocities
represents the gas drawn from the $-3$ and/or $9$ km s$^{-1}$ clouds.
Similar feature has been found also in other star forming regions
where cloud-cloud collisions are suggested
\citep[e.g.,][]{Dobashi2019}.

We should note that the $^{12}$CO emission with the intermediate velocities
could be partially due to
the outflows driven by DR21 and DR21(OH). However, the outflow lobes are not overlapping
with the peak position of the $-9$ km s$^{-1}$ cloud but they are extending toward other directions
(figure \ref{fig:outflows}), and thus the emission with the intermediate velocities just around the peak position
of the $-9$ km s$^{-1}$ cloud
is unlikely due to the outflows. In addition, as shown in figure \ref{fig:intermediate},
distribution of the $^{12}$CO emission with the intermediate velocities (in the range $1<V_{\rm LSR}<5$ km s$^{-1}$)
delineates the round edge of the densest part of the 9 km s$^{-1}$ cloud (see the white ellipse pointed
by an arrow in the figure), though the emission just around DR21 is apparently due to the outflow.
This indicates the physical relation of the 9 km s$^{-1}$ cloud and the intermediate velocity component,
providing us with an additional support to suggest 
that the emission represents the gas drawn from
the $-3$ and/or $9$ km s$^{-1}$ clouds.
 
In Sgr B2 where \citet{Hasegawa1994} discovered a clear evidence of the cloud-cloud collisions
for the first time,
they identified a clump at $V_{\rm LSR}=70-80$ km s$^{-1}$ which nicely fit to
a hole in the cloud complex at $V_{\rm LSR}=40-50$ km s$^{-1}$, and young massive stars
(including compact H{\scriptsize{II}} regions and maser sources) are aligned along the interface
of the clump and hole. This picture appears similar to the case of the DR21 cloud:
As seen in figure \ref{fig:collision}, the two massive YSOs, DR21 and DR21(OH), are located
in the periphery of the dense clump in the $9$ km s$^{-1}$ cloud.
A question which immediately arises is that why they are not located at the very center
of the $9$ km s$^{-1}$ clump which should be the center of the collision.
Recent numerical simulations show that, when the time comparable to the free-fall time
has passed after the collision of two clouds,
a dense layer consisting of a number of clumps/cores are formed along an arc (or parabolic surface)
around the axis of the collision
\citep[see figure 13 of][]{Takahira2018}. Massive stars should form there.
This may be the case for the DR21 and DR21(OH) as well as for the YSOs in Sgr B2.

To conclude, we suggest that the $-3$ and $9$ km s$^{-1}$ clouds are colliding against
each other and that the formation of the massive YSOs,
i.e., DR21, DR21(OH), W75S-FIR1, W75S-FIR2, and W75S-FIR3,
were likely to be triggered by the collision.
We summarize our picture of this region in a schematic
illustration in figure \ref{fig:model}.
Our results strongly support the hypothesis of \citet{Dickel1978}
that the two clouds are interacting with each other.

As summarized in table \ref{tab:mass}, the natal cloud of DR21 (the $-3$ km s$^{-1}$ cloud)
has a total mass and mean molecular column density of $\sim 3 \times 10^4$ $M_\odot$ and $\sim3 \times 10^{23}$ cm$^{-2}$, respectively.
It is colliding with the $-9$ km s$^{-1}$ cloud at a relative velocity of $\sim12$ km s$^{-1}$,
and a cluster containing several O-type stars are forming in DR21 \citep[e.g.,][]{Harris1973}.
We compare these features of DR21 with other clouds producing O-type stars
whose formation is supposed to be triggered by cloud-cloud collisions.
\citet{Fukui2018} recently summarized properties of such clouds in the literature.
Looking into their table 1, relative radial velocities of the colliding clouds are
$\sim10-20$ km s$^{-1}$, and except for a few cases, the clouds, not necessarily both of them,
have a mass of $\sim10^{4}-10^5$ $M_\odot$.
Though the sample collected by \citet{Fukui2018} is rather small,
parameters of the DR21 cloud are in these ranges, suggesting that DR21 may be a typical
case of massive star formation induced by cloud-cloud collisions.
In addition, \citet{Fukui2018} suggested that the clouds produce
so-called ``super cluster" containing $\sim10-20$ O-type stars
if the column densities of the clouds are as high as $N({\rm H_2})\simeq10^{23}$ cm$^{-2}$, 
whereas less dense clouds
($\sim10^{22}$ cm$^{-2}$) produce only a single O-type star. 
DR21 apparently corresponds to the former case, and is similar to RCW 38
in table 1 of \citet{Fukui2018}
forming a young cluster with an age of $\sim0.1$ Myr.

Finally, cloud-cloud collisions
like that found in the DR21 cloud give us a hint to
understand the onset of formation of massive stars and star clusters.
Recent studies of young clusters have shown that they are forming
in massive clumps with a typical mass of $\sim 1 \times 10^3$ $M_\odot$
\citep[e.g.,][]{Saito2007,Shimoikura2013}. Statistical studies carried out by \citet{Shimoikura2018}
have revealed that most of the clumps associated with very young clusters
are gravitationally collapsing, exhibiting a clear clump-scale infalling motion with rotation
\citep[see also][]{Shimoikura2016}. Other clumps not forming clusters are
expected to be in a stage prior to cluster formation and to be younger
than those forming clusters. However, \citet{Shimoikura2018} found that
the clumps without clusters are more evolved in terms of chemical compositions
than the ones already forming clusters. They suggested that the clumps are gravitationally
stable having survived for a long time without collapsing due to cloud-supporting forces, e.g.,
by the magnetic field. Cloud-cloud collision should cause distortion
of the magnetic field as well as sudden enhancement of gas density by compression,
which should force the cloud to collapse \citep[e.g.,][]{Wu2017a,Wu2017b}.
In the DR21 cloud, the strong magnetic field of an order of $\gtrsim1$ mG is
actually inferred \citep[e.g.,][]{Lai2003,Ching2018}
and is also suggested to be interacting with the gas dynamics \citep[][]{Ching2018}.
The DR21 cloud is therefore very likely to be
experiencing such a collision-induced star formation at present.


\section{Conclusions} \label{sec:conclusions}

We performed molecular observations of the DR21 region
using the 45-m telescope at the Nobeyama Radio Observatory (NRO).
We revealed the global distributions of the molecular lines,
and identified distinct clouds in this region.
Main conclusions of this paper are summarized in the following points.

\begin{enumerate}

\item We mapped an area of $\sim 8\arcmin \times 12\arcmin$ around the DR21 region
with twenty molecular lines including the $^{12}$CO($J=1-0$), $^{13}$CO($J=1-0$), HCO$^{+}$($J=1-0$), and SiO($J=2-1, v=0$) emission lines.
Among them, sixteen lines were significantly detected.
Based on the $^{12}$CO and $^{13}$CO data, we identified
five velocity components at
the radial velocities $V_{\rm LSR} \simeq -42$, $-22$, $-3$, $9$, and $17$ km s$^{-1}$.
We call clouds associated with these components ``$-42$", ``$-22$",``$-3$", ``$9$", and ``$17$" km s$^{-1}$ clouds.

\item Based on the $^{12}$CO and $^{13}$CO data, we estimated the total molecular masses
of the $-3$, $9$, and $17$ km s$^{-1}$ clouds within the observed area to be
$\sim31000$, $\sim3400$, and $\sim500$ $M_\odot$, respectively, assuming the local thermodynamic equilibrium (LTE).
We found that there is a clear anticorrelation between the $-3$ and $9$ km s$^{-1}$ clouds,
suggesting that
they are colliding against each other.

\item The SiO($J=2-1,v=0$) emission at $86.8$ GHz was detected around young stellar objects
in the DR21 region. Distributions of the emission line are well correlated with those of molecular outflows
traced in the $^{12}$CO and a few other emission lines.

\item
Around the intersection of the $-3$ and $9$ km s$^{-1}$ clouds, we found a velocity component having
intermediate velocities and bridging the two clouds. We also found that the geometrical relation of the
YSOs and the intersection is consistent with the results of recent numerical simulations of cloud-cloud collisions.
These findings indicate that the $-3$ and $9$ km s$^{-1}$
clouds are colliding, and that the collision
induced the formation of the massive stars in the DR21 cloud. 
The interaction of the $-3$ and $9$ km s$^{-1}$ clouds in this region was first
suggested by \citet{Dickel1978}, and our results strongly support their hypothesis.

\end{enumerate}


\begin{ack}
This work was financially supported by JSPS KAKENHI Grant Numbers
JP17H02863, JP17H01118, JP26287030, and JP17K00963. 
The 45-m radio telescope is operated by NRO, a branch of NAOJ. 
YS received support from the ANR (project NIKA2SKY, grant agreement
ANR-15-CE31-0017).
\end{ack}

\appendix
\section{Maps of the other molecular emission lines}\label{sec:appendix}
\begin{figure*}
\begin{center}
\includegraphics[scale=0.3]{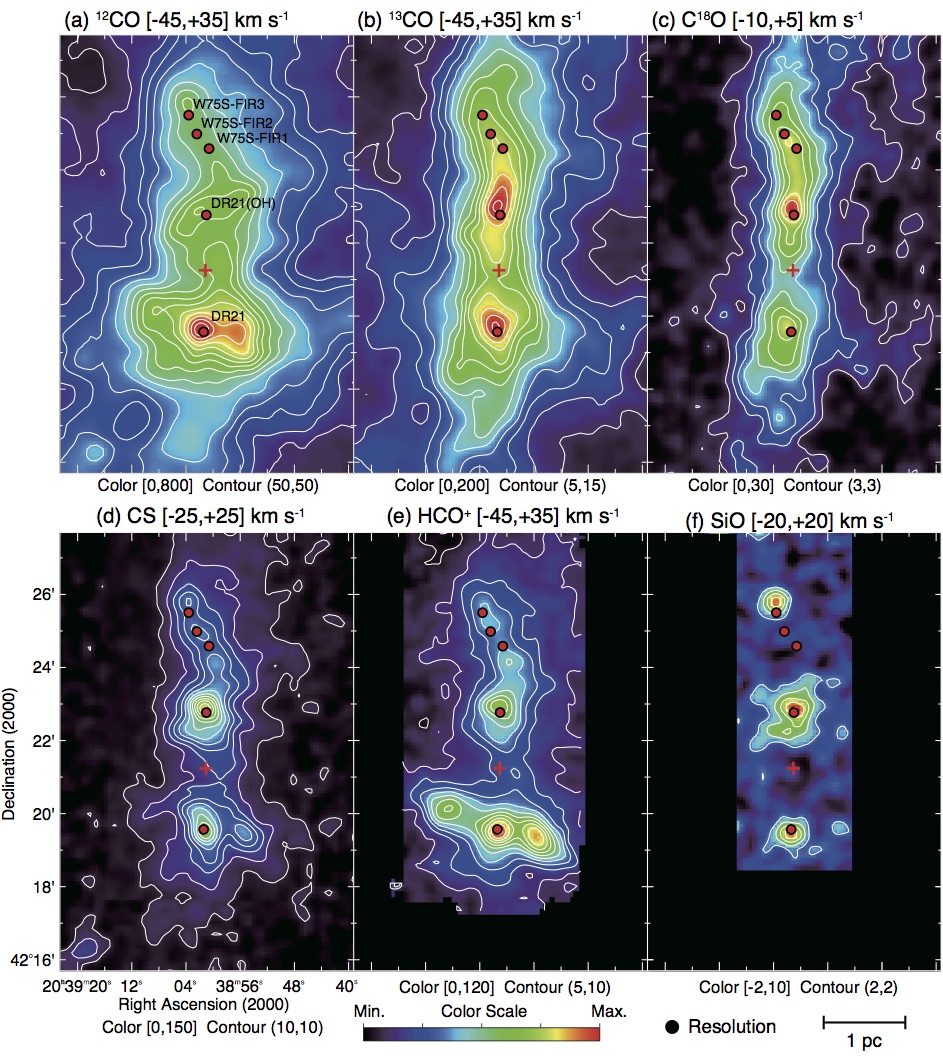}
\end{center}
\caption{
Velocity-integrated intensity maps of the observed emission lines.
Molecular lines and LSR velocity ranges used for the integration
are denoted above each panel.
A set of the minimum and maximum values of the color scale
is given in the brackets, and a set of the lowest contours and the contour intervals
is given in the parentheses below each panel (in units of K km s$^{-1}$).
Red circles denote positions of YSOs, and plus signs denote the intensity peak position of the 9 km s$^{-1}$ cloud.
Areas observed in HCO$^+$ and SiO are smaller than
those observed in the other molecular lines.
Filled circle below panel (f) denotes the angular resolution of the maps ($23\arcsec$).
Molecular lines not discussed in the main text are shown in the appendix (figure \ref{fig:iimap2}).
\label{fig:iimap1}}
\end{figure*}

\begin{figure*}
\begin{center}
\includegraphics[scale=0.3]{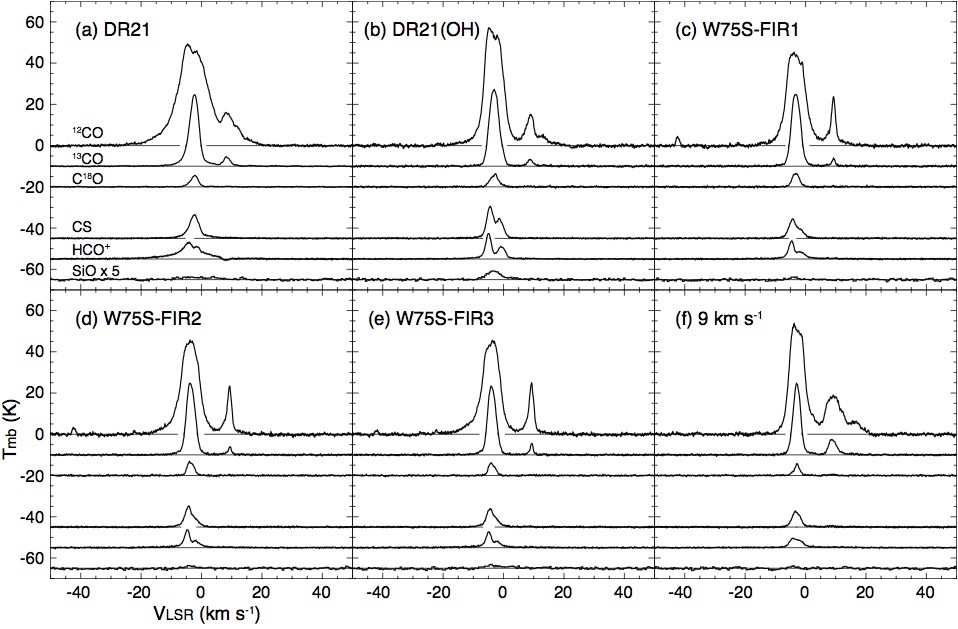}
\end{center}
\caption{
The $^{12}$CO, $^{13}$CO, C$^{18}$O, CS, HCO$^+$,
and SiO spectra
observed toward the five YSOs as well as toward the intensity-peak position of
the 9 km s$^{-1}$ cloud.
The $^{13}$CO, C$^{18}$O, CS, HCO$^+$, and SiO spectra are offset
by $-10$, $-20$, $-45$, $-55$, and $-65$ K, respectively.
The SiO spectra are scaled up by a factor of 5.
\label{fig:spectra}}
\end{figure*}

\begin{figure*}
\begin{center}
\includegraphics[scale=0.3]{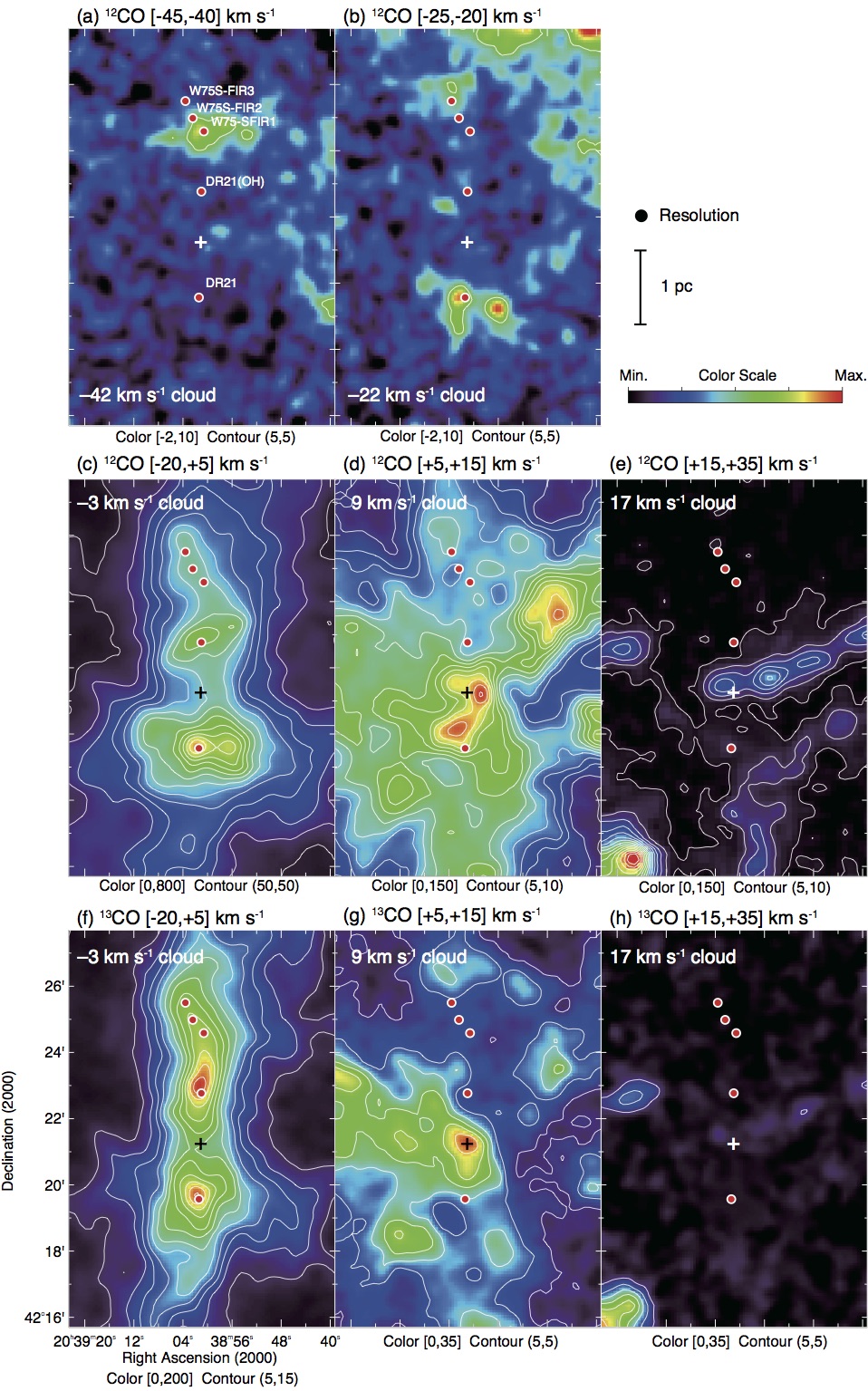}
\end{center}
\caption{
Distributions of the $-42$, $-22$, $-3$, $9$, and $17$ km s$^{-1}$ clouds.
Panels (a)--(e) are the channel maps of the $^{12}$CO emission line, and
panels (f)--(h) are those of the $^{13}$CO emission line
which are made for the velocity intervals same as for (c)--(e) in this order.
Velocity intervals used for the integration are denoted above each panel.
Others are the same as in figure \ref{fig:iimap1}.
\label{fig:channel_maps}}
\end{figure*}

\begin{figure}
\begin{center}
\includegraphics[scale=0.3]{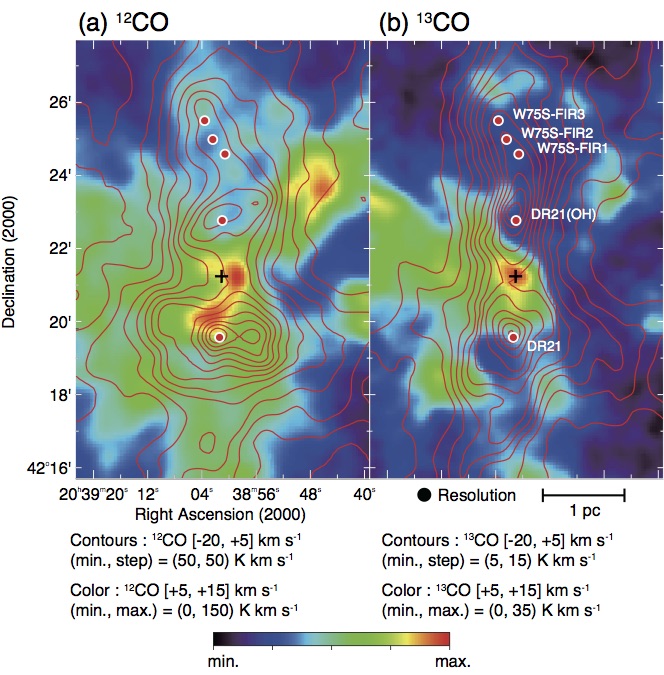}
\end{center}
\caption{
The (a) $^{12}$CO and (b) $^{13}$CO distributions of the $-3$ km s$^{-1}$ cloud (contours)
and the 9 km s$^{-1}$ cloud (color) which are the same as those
in panels (c), (d), (f), and (g) of figures \ref{fig:channel_maps}.
Plus signs denote the intensity peak position of the 9 km s$^{-1}$ cloud in $^{13}$CO.
\label{fig:collision}}
\end{figure}

\begin{figure*}
\begin{center}
\includegraphics[scale=0.3]{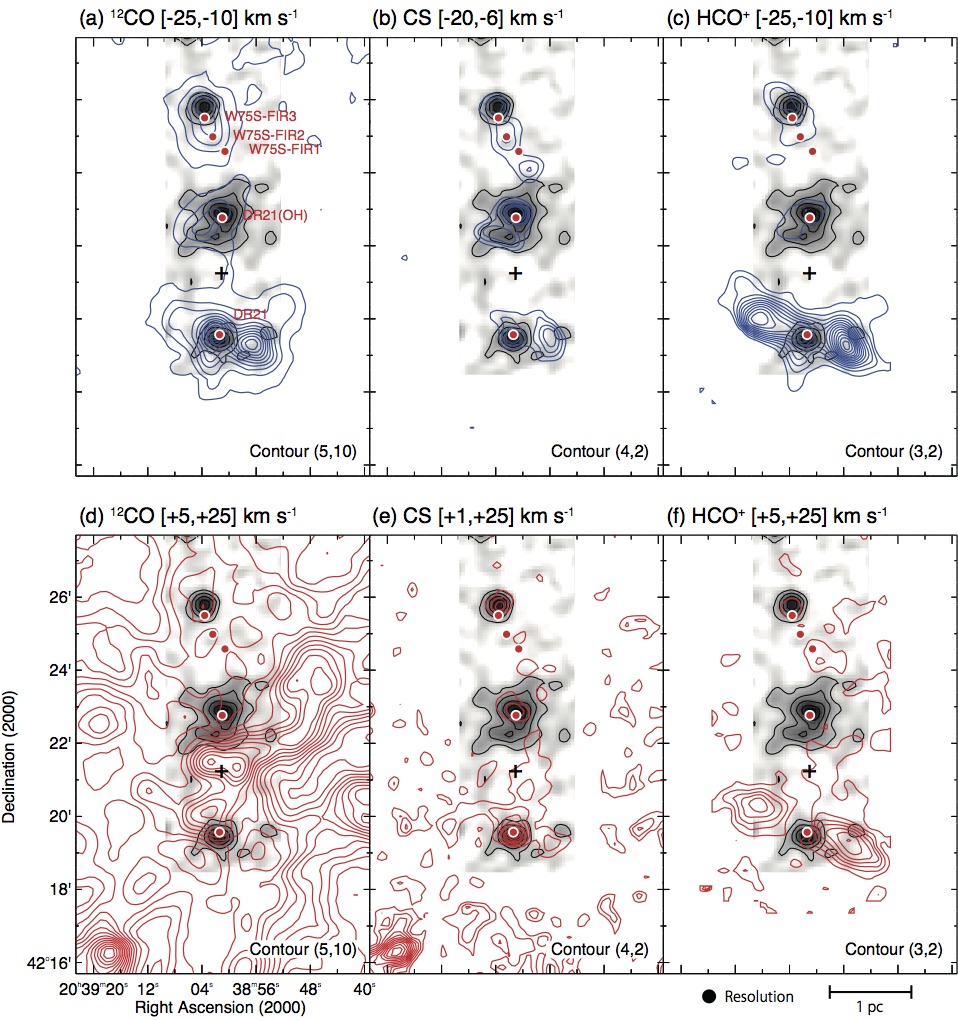}
\end{center}
\caption{
Distributions of molecular outflows detected in $^{12}$CO, CS, and HCO$^+$ (blue and red contours)
overlaid with the SiO intensity (gray scale and black contours).
Blue lobes are shown in panels (a)--(c), and red lobes are shown in panels (d)-(f).
Velocity ranges used for the integration are denoted above each panel,
and the lowest contours and contour intervals for the outflow lobes are indicated in the parentheses
in the bottom-right corner of each panel.
Contours for the SiO intensity are the same as in figure \ref{fig:iimap1}f.
Red lobes in panels (d) are highly contaminated by other velocity components.
\label{fig:outflows}}
\end{figure*}

\begin{figure}
\begin{center}
\includegraphics[scale=0.25]{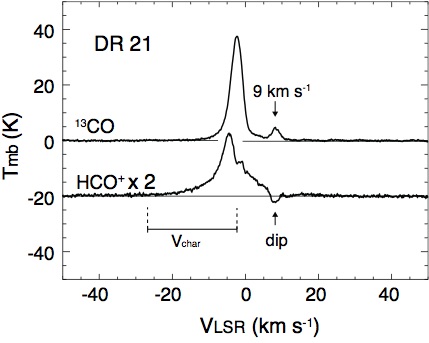}
\end{center}
\caption{
The $^{13}$CO and HCO$^{+}$ spectra observed toward DR21, same as those shown in figure \ref{fig:spectra}a.
There is a dip in the HCO$^{+}$ spectrum at the velocity of the $9$ km s$^{-1}$ cloud.
The HCO$^{+}$ spectrum is scaled up by a factor of 2, and is offset by $-20$ K.
$V_{\rm char}$ is the characteristic velocity of the outflow defined as the maximum velocity separation
of the high velocity wing(s) from the systemic velocity of DR21
(about $-2.4$ km s$^{-1}$, see the radial velocities of $^{13}$CO and C$^{18}$O in table \ref{tab:line_parameters}).
\label{fig:hco}}
\end{figure}

\begin{figure}
\begin{center}
\includegraphics[scale=0.3]{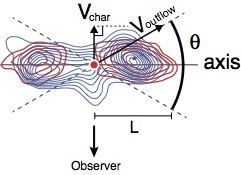}
\end{center}
\caption{
Schematic illustration of the outflow associated with DR21 plotted by the red filled circle (see section \ref{sec:outflow}).
\label{fig:outflow_age}}
\end{figure}

\begin{figure*}
\begin{center}
\includegraphics[scale=0.3]{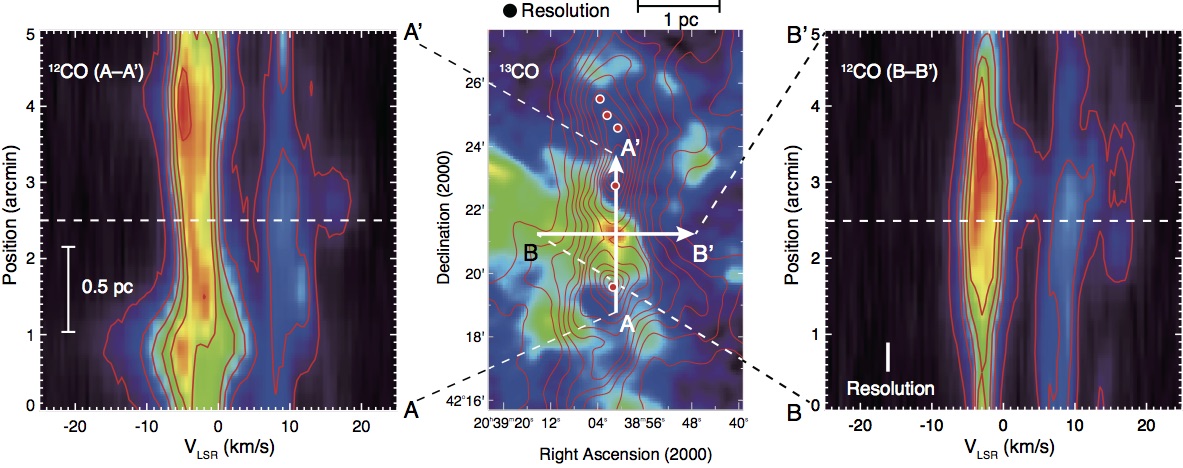}
\end{center}
\caption{
Position-velocity diagrams measured in $^{12}$CO along the cut A--A' (left panel) and B--B' (right panel)
denoted in the middle panel. Contours start from 5 K with a step of 10 K. The horizontal white broken lines
denote the position of the intensity peak position of the 9 km s$^{-1}$ cloud.
In the diagrams, the $^{12}$CO spectra are smoothed to the 0.5 km s$^{-1}$ velocity resolution
as denoted in the right panel.
\label{fig:pv}}
\end{figure*}

\begin{figure}
\begin{center}
\includegraphics[scale=0.3]{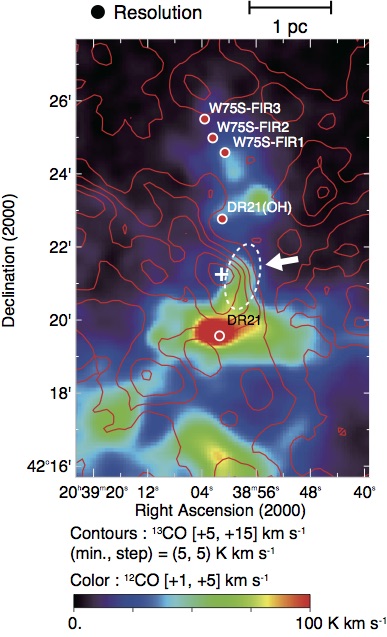}
\end{center}
\caption{
Intensity distribution of the $^{12}$CO emission line integrated over the
velocity range $1 <V_{\rm LSR}<5$ km s$^{-1}$ (color).
The $^{13}$CO intensity (contours) of the 9 km s$^{-1}$ cloud, same as
figure \ref{fig:channel_maps}g, is superposed for comparison.
The $^{12}$CO emission in the map originates mostly from the molecular outflows,
but the emission denoted by the white broken ellipse with an arrow
delineates the round edge of the 9 km s$^{-1}$ cloud and is likely to represent the
intermediate velocity components generated by the collision.
\label{fig:intermediate}}
\end{figure}

\begin{figure}
\begin{center}
\includegraphics[scale=0.3]{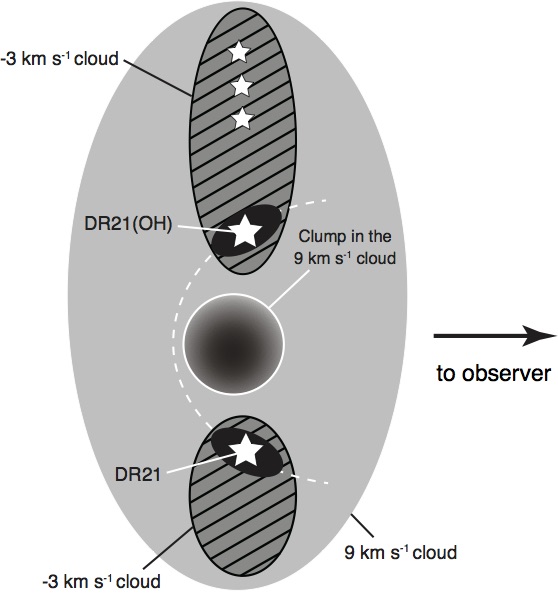}
\end{center}
\caption{
Schematic illustration of the $-3$ and $9$ km s$^{-1}$ clouds discussed in section \ref{sec:discussion}.
The hatched regions denote the $-3$ km s$^{-1}$ cloud.
The dense part of the $9$ km s$^{-1}$ cloud is shown by the gray circle with white line.
Star symbols represent the YSOs same as in the other figures.
Black ellipses around DR21 and DR21(OH) are the dense clumps
forming along the arc (or parabolic surface) denoted by the white broken line (see text).
\label{fig:model}}
\end{figure}

\begin{figure*}
\begin{center}
\includegraphics[scale=0.3]{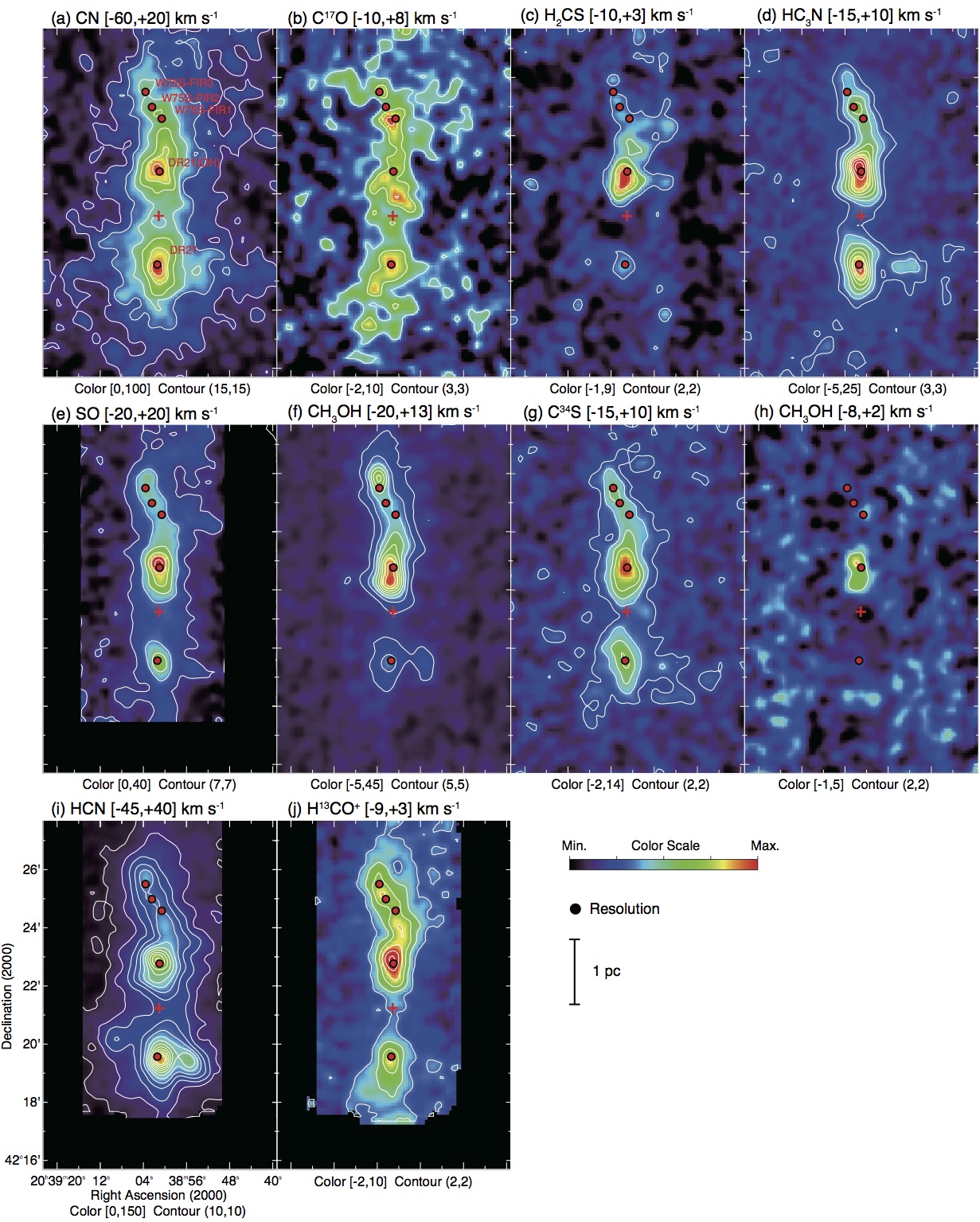}
\end{center}
\caption{
Same as figure \ref{fig:iimap1}, but for molecular emission lines not discussed
in the main text.
Areas observed in SO, HCN, and H$^{13}$CO$^{+}$ are smaller than
those observed in the other molecular lines.
For the CN, C$^{17}$O, and HCN emission lines, all of the hyperfine lines are integrated.
\label{fig:iimap2}}
\end{figure*}

In figure \ref{fig:iimap2}, we display the velocity-integrated intensity maps of
the ten molecular emission lines not shown in figure \ref{fig:iimap1}.




\clearpage



\begin{table*}
\caption{Observed molecular lines} 
\begin{tabular}{cccccc}  
\hline\noalign{\vskip3pt} 
	\multicolumn{1}{c}{Molecule} & \multicolumn{1}{c}{Transition} & \multicolumn{1}{c}{Rest Frequency} & \multicolumn{1}{c}{$\Delta T_{\rm mb}$} &   \multicolumn{1}{c}{Detection} &   \multicolumn{1}{c}{Figure}\\
	\multicolumn{1}{c}{} & \multicolumn{1}{c}{} & \multicolumn{1}{c}{(GHz)}  & \multicolumn{1}{c}{(K)} &  \multicolumn{1}{c}{} &  \multicolumn{1}{c}{Number}\\  [2pt] 
\hline\noalign{\vskip3pt} 
$^{12}$CO		& $J=1-0$				&	115.271202	& 	1.4		& Y  & \ref{fig:iimap1}a \\ %
CN				& $J=3/2-1/2$			&	113.490982	&	1.3		& Y  & \ref{fig:iimap2}a  \\ %
CCS				& $N,J=9,8-8,7$		&	113.410204	&	1.5		& N  & ...  \\ %
C$^{17}$O		& $J=1-0$, $F=7/2-5/2$	&	112.358988	&	1.0		& Y  & \ref{fig:iimap2}b  \\ %
CH$_3$CN		& $6(2)-5(2),$ $F=7-6$	&	110.375052	& 	0.6		& N  & ... \\ %
$^{13}$CO		& $J=1-0$				&	110.201354	& 	0.7		& Y  & \ref{fig:iimap1}b \\ %
NH$_2$D			& $1(1,1)0-1(0,1)0+$		&	110.153599	&	1.0		& N  & ...  \\ %
C$^{18}$O		& $J=1-0$				&	109.782173	& 	0.6		& Y  & \ref{fig:iimap1}c \\  %
H$_2$CS			& $3(1,3)-2(1,2)$		&	101.477885	&	0.6		& Y  & \ref{fig:iimap2}c  \\ %
HC$_3$N			& $J=11-10$			&	100.076386	&	0.5		& Y  & \ref{fig:iimap2}d  \\ %
SO				& $N,J=5,4-4,4$		&	100.029565	&	0.3		& N  & ...  \\%
SO				& $N,J=2,3-1,2$		&	  99.299905	&	0.3		& Y  & \ref{fig:iimap2}e  \\%
CS				& $J=2-1$				&	  97.980953	&	0.3		& Y  & \ref{fig:iimap1}d  \\ %
CH$_3$OH		& $2(0,2)-1(0,1)A++$	&	  96.741377	&	0.5		& Y  & \ref{fig:iimap2}f  \\ %
C$^{34}$S		& $J=2-1$				&	  96.412950	&	0.4		& Y  & \ref{fig:iimap2}g  \\ %
CH$_3$OH		& $2(1,2)-1(1,1)A++$	&	  95.914310	&	0.5		& Y  & \ref{fig:iimap2}h  \\ %
HCO$^+$			& $J=1-0$				&	  89.188526	& 	0.3		& Y  & \ref{fig:iimap1}e \\ %
HCN				& $J=1-0$, $F=2-1$		&	 88.6318473	&	0.2		& Y  & \ref{fig:iimap2}i  \\ %
SiO				& $J=2-1$, $v=0$		&	  86.846995	& 	0.2		& Y  & \ref{fig:iimap1}f \\ %
H$^{13}$CO$^+$	& $J=1-0$				&	  86.754288	&	0.2		& Y  & \ref{fig:iimap2}j  \\ %
\hline
\end{tabular} \label{tab:lines}
   \begin{tabnote}
The rest frequencies are taken from Splatalogue.
$\Delta T_{\rm mb}$ is the 1 $\sigma$ noise level measured for a velocity
resolution of 0.1 km s$^{-1}$.
`Y' and `N' in the fifth column mean detection and non-detection, respectively.
`Figure Number' in the last column denotes the figure numbers where 
intensity distributions of the detected lines are displayed.
   \end{tabnote}
\end{table*} 
\clearpage


\begin{table*}
\caption{Gaussian parameters of the CO lines} 
\scriptsize
\begin{tabular}{cccccccccccc}  
\hline\noalign{\vskip3pt} 
	\multicolumn{1}{c}{Position} & \multicolumn{1}{c}{R.A. (J2000)} & \multicolumn{1}{c}{Dec. (J2000)}& \multicolumn{1}{c}{$T^{^{12}{\rm CO}}_{\rm mb}$} & \multicolumn{1}{c}{$V^{^{12}{\rm CO}}_{\rm LSR}$} & \multicolumn{1}{c}{$\Delta V^{^{12}{\rm CO}}$} & \multicolumn{1}{c}{$T^{^{13}{\rm CO}}_{\rm mb}$} & \multicolumn{1}{c}{$V^{^{13}{\rm CO}}_{\rm LSR}$} & \multicolumn{1}{c}{$\Delta V^{^{13}{\rm CO}}$} & \multicolumn{1}{c}{$T^{{\rm C^{18}O}}_{\rm mb}$} & \multicolumn{1}{c}{$V^{{\rm C^{18}O}}_{\rm LSR}$} & \multicolumn{1}{c}{$\Delta V^{{\rm C^{18}O}}$} \\
	\multicolumn{1}{c}{}& \multicolumn{1}{c}{($^{\rm h}$ $^{\rm m}$ $^{\rm s}$)} & \multicolumn{1}{c}{($^\circ$ $\arcmin$ $\arcsec$)} & \multicolumn{1}{c}{(K)} & \multicolumn{1}{c}{(km s$^{-1}$)} & \multicolumn{1}{c}{(km s$^{-1}$)} & \multicolumn{1}{c}{(K)} & \multicolumn{1}{c}{(km s$^{-1}$)} & \multicolumn{1}{c}{(km s$^{-1}$)}  & \multicolumn{1}{c}{(K)} & \multicolumn{1}{c}{(km s$^{-1}$)} & \multicolumn{1}{c}{(km s$^{-1}$)} \\  [2pt] 
\hline\noalign{\vskip3pt} 
DR21	&20 39 01.4&42 19 34     &  $48.4$ & $-3.0$ & $11.3$ & $34.1$ & $-2.5$ & $3.8$ & $5.3$ & $-2.4$ & $3.1$ \\
		&                 &                   &  $12.4$ & $ 9.7$ & $ 6.7$ & $ 4.1$ & $ 8.3$ & $3.2$ &  ...  &  ...   & ...   \\
\hline
DR21(OH)&20 39 01.0&42 22 46    &  $59.3$ & $-3.3$ & $ 6.8$ & $37.8$ & $-3.1$ & $3.8$ & $5.7$ & $-3.0$ & $3.3$ \\
		&                 &                  &  $12.5$ & $ 9.0$ & $ 4.8$ & $ 3.1$ & $ 8.8$ & $2.5$ &  ...  &  ...   & ...   \\
\hline
W75S-FIR1&20 39 00.6&42 24 35  & $46.7$ & $-3.2$ & $ 7.1$ & $36.0$ & $-3.2$ & $3.5$ & $6.6$ & $-3.2$ & $2.8$ \\
		&                 &                  &  $21.0$ & $ 9.2$ & $ 2.1$ & $ 3.4$ & $ 9.3$ & $1.6$ &  ...  &  ...   & ...   \\
\hline
W75S-FIR2&20 39 02.4&42 24 59  & $46.2$ & $-3.8$ & $ 7.0$ & $35.2$ & $-3.6$ & $3.4$ & $6.6$ & $-3.6$ & $2.7$ \\
		&                    &               &  $21.0$ & $ 9.2$ & $ 2.2$ & $ 3.4$ & $ 9.3$ & $1.6$ &  ...  &  ...   & ...   \\
\hline
W75S-FIR3&20 39 03.6&42 25 30 &  $46.1$ & $-4.1$ & $ 6.6$ & $33.7$ & $-3.9$ & $3.2$ & $5.9$ & $-3.9$ & $2.7$ \\
		&                    &                 &  $22.7$ & $ 9.2$ & $ 2.2$ & $ 5.3$ & $ 9.3$ & $1.3$ &  ...  &  ...   & ...   \\
\hline
9 km s$^{-1}$&20 39 01.0&42 21 17&  $55.1$ & $-2.8$ & $ 5.7$ & $34.5$ & $-2.9$ & $2.9$ & $5.1$ & $-2.8$ & $2.3$ \\
 cloud		&              &              & $18.2$ & $ 9.2$ & $ 6.7$ & $ 7.3$ & $ 8.9$ & $3.2$ &  ...  &  ...   & ...   \\
			&              &              &  $ 5.1$ & $16.9$ & $ 4.4$ &  ...   &  ...  &  ...  &  ...  &  ...   & ...   \\ 
\hline
\end{tabular} \label{tab:line_parameters}
   \begin{tabnote}
The table lists the Gaussian parameters of the velocity components seen in the $^{12}$CO, $^{13}$CO, and C$^{18}$O spectra in figure \ref{fig:spectra}.
Coordinates of the YSOs are taken from \citet{Motte2007}:
DR21, DR21(OH), W75S-FIR1, W75S-FIR2, and W75S-FIR3
correspond to N46, N44, N43, N51, and N54 in their table 1, respectively.
The peak brightness temperature $T_{\rm mb}$, the centroid velocity $V_{\rm LSR}$, and the line width $\Delta V$ (FWHM)
are measured by fitting the observed spectra with a Gaussian function with 1--3 components.
   \end{tabnote}
\end{table*} 

\clearpage

\begin{table}
\caption{Properties of the clouds} 
\begin{tabular}{cccccr}  
\hline\noalign{\vskip3pt} 
	\multicolumn{1}{c}{Cloud} & \multicolumn{1}{c}{$T_{\rm ex}^{\rm max}$}  & \multicolumn{1}{c}{$N^{\rm max}$(H$_2$)} & \multicolumn{1}{c}{$\overline{N}$(H$_2$)} & \multicolumn{1}{c}{$S$}& \multicolumn{1}{c}{$M$} \\
	\multicolumn{1}{c}{} & \multicolumn{1}{c}{(K)}  & \multicolumn{1}{c}{(cm$^{-2}$)} & \multicolumn{1}{c}{(cm$^{-2}$)} & \multicolumn{1}{c}{(pc$^2$)} & \multicolumn{1}{c}{($M_\odot$)} \\  [2pt] 
\hline\noalign{\vskip3pt} 
$-3$ km s$^{-1}$	&	$69.5$	& $4.72 \times 10^{23}$ & $2.99 \times 10^{23}$ & 1.29 & $31,080$ 	\\
$9$ km s$^{-1}$	&	$32.3$	& $2.90 \times 10^{22}$ &	 $1.82 \times 10^{22}$ & 2.56 & $3,370$ 	\\
$17$ km s$^{-1}$	&	$37.1$	& $3.41 \times 10^{22}$ &	 $2.83 \times 10^{22}$ & 0.22 & $480$  \\
\hline
\end{tabular} \label{tab:mass}
   \begin{tabnote}
The table lists
$T_{\rm ex}^{\rm max}$ the maximum excitation temperature,
$N^{\rm max}$(H$_2$) the maximum H$_2$ column density,
$\overline{N}$(H$_2$) the mean H$_2$ column density,
$S$ the surface area defined at one half of the $N^{\rm max}$(H$_2$) level,
and
$M$ the total mass
of the $-3$, $9$, and $17$ km s$^{-1}$ clouds within the region displayed in figure \ref{fig:channel_maps}.
$\overline{N}$(H$_2$) is the average value within $S$.
 \end{tabnote}
\end{table} 
\clearpage



\end{document}